\documentclass[journal,usenames,dvipsnames]{IEEEtran}
\IEEEoverridecommandlockouts

\usepackage{ifpdf}
\ifpdf
   \usepackage{url} 
\else
\fi

\usepackage{mystyle}

\begin{document}

\title{In-network Congestion-aware Load Balancing\\at Transport Layer}

\author{\IEEEauthorblockN{
    Ashkan Aghdai\IEEEauthorrefmark{1},
    Michael I.-C. Wang\IEEEauthorrefmark{1}\IEEEauthorrefmark{3},
    Yang Xu\IEEEauthorrefmark{2},
    Charles H.-P. Wen\IEEEauthorrefmark{3},
    H. Jonathan Chao\IEEEauthorrefmark{1}}
\IEEEauthorblockA{
    \\ \IEEEauthorrefmark{1}Tandon School of Engineering, New York University, Brooklyn, NY, USA
    }
\and
\IEEEauthorblockA{
    \\ \IEEEauthorrefmark{2}School of Computer Science, Fudan University, Shanghai, China
    }
\and
\IEEEauthorblockA{
    \\ \IEEEauthorrefmark{3}Department of Electrical and Computer Engineering, National Chiao Tung University, Taiwan
    }
}

\newcommand\copyrighttext{%
  \footnotesize This article is submitted to 2019 IEEE conference on NFV/SDN and is under review.}
\newcommand\copyrightnotice{%
\begin{tikzpicture}[remember picture,overlay]
\node[anchor=south,yshift=10pt] at (current page.south) {\fbox{\parbox{\dimexpr\textwidth-\fboxsep-\fboxrule\relax}{\copyrighttext}}};
\end{tikzpicture}%
}

\maketitle
\copyrightnotice

\begin{abstract}
    Load balancing at transport layer is an important function in data centers, content delivery networks, and mobile networks, where per-connection consistency (PCC) has to be met for optimal performance.
    Cloud-native L4 load balancers are commonly deployed as virtual network functions (VNFs) and are a critical forwarding element in modern cloud infrastructure.
    We identify load imbalance among service instances as the main cause of additional processing delay caused by transport-layer load balancers.
    Existing transport-layer load balancers rely on one of two methods: host-level traffic redirection, which may add as much as 12.48\% additional traffic to underlying networks, or connection tracking, which consumes a considerable amount of memory in load balancers.
    Both of these methods result in inefficient usage of network resources.
    
    We propose the in-network congestion-aware load Balancer (INCAB) to achieve even load distribution among service instances and optimal network resources usage in addition to meeting the PCC requirement.
    We show that INCAB is capable of identifying and monitoring each instance's most-utilized resource and can improve the load distribution among all service instances.
    INCAB utilizes a Bloom filter and an ultra-compact connection table for in-network flow distribution.
    Furthermore, it does not rely on end hosts for traffic redirection.
    Our flow level simulations show that INCAB improves flows' average completion time by 31.97\% compared to stateless solutions.
\end{abstract}


\section{Introduction}
Transport-layer load balancing is a critical function in data centers \cite{patel2013ananta,eisenbud2016maglev,miao2017silkroad,olteanu2018stateless}, content delivery networks (CDNs)~\cite{araujo2018balancing}, and mobile networks~\cite{aghdai2018transparent,aghdai2019enabling}.
Efficient load balancing results in even distribution of load among serving instances which in turn leads to reduced average flow completion time and improved end-to-end latency for users.
Cloud-native L4 load balancer VNFs are usually deployed in close proximity of the service instances using software instances~\cite{patel2013ananta,eisenbud2016maglev,gandhi2015duet}, programmable switches~\cite{miao2017silkroad}, or programmable network interface cards~\cite{aghdai2018spotlight}.
These VNFs are one of the most important forwarding elements in cloud infrastructure due to their impact on users' experience.

Research on L4 load balancing has been very active during recent years, but the focus has always been on load balancers' most basic function in providing per-connection consistency (PCC).
Very few proposals define performance metrics such as aggregated service throughput~\cite{aghdai2018spotlight} or fairness~\cite{yoann20186lb,claudel2018stateless} among service instances and the solutions are either too complicated or require a lot of network resources.

Load imbalance could happen due to a variety of reasons, chief among which is the imbalance in input traffic.
Existing works utilize equal-cost multipath routing (ECMP) or various forms of consistent hashing~\cite{karger1997consistent} to roughly distribute equal number of connections among service instances.
However, not only incoming connections' size has a heavy-tailed distribution~\cite{benson2010network,greenberg2009vl2,aghdai2013traffic}, but also instances can have different capacities when they were deployed incrementally over time.
Hence, some service instances that receive elephant flows or have smaller capacity can be overloaded while the rest may be under-utilized.
As a result, users' experience will vary depending on the serving machine and its load.

State-of-the-art load balancers such as Beamer~\cite{olteanu2018stateless}, Faild~\cite{araujo2018balancing}, and SHELL~\cite{claudel2018stateless}, rely on end-host traffic redirection to meet PCC.
This method alleviates the pressure on load balancers by offloading connection tracking to hosts.
However, a portion of the traffic is redirected back to the data center network.
The additional traffic caused by the rerouting may be detrimental to the operation of data center network, especially if the network is over-subscribed.
Beamer and Faild trigger traffic rerouting when an instance is added to or removed from the service pool and the amount of rerouted traffic is minimal, whereas in SHELL, rerouting may also happen if the load balancer sends new connections to an overwhelmed service instance.
As a result, traffic rerouting occurs at a much higher frequency, resulting in additional traffic rerouting in SHELL.
Silkroad~\cite{miao2017silkroad} and other solutions that rely on connection tracking~\cite{gandhi2015duet,aghdai2018spotlight} use significant amount of memory at load balancers.

\begin{figure}[t]
    \centering
    \includegraphics[width=0.4\textwidth]{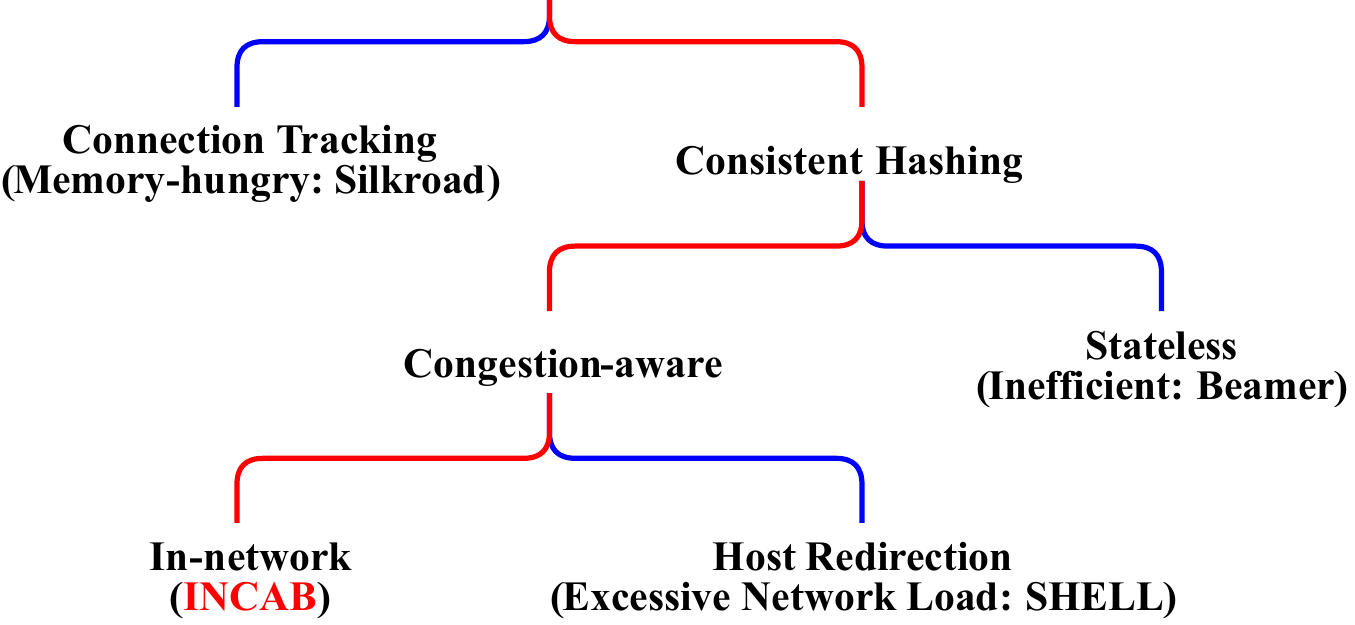}
    \caption{Load balancers' classification}
    \label{fig0}
\end{figure}

We propose the in-network congestion-aware load balancer (INCAB) to optimize the load distribution without relying on hosts to reroute any traffic.
Our solution relies on service instances to notify load balancers if they experience congestion; load balancers in turn, prioritize sending traffic to uncongested hosts without relying on connection tracking or traffic rerouting.
INCAB utilizes a Bloom filter~\cite{bloom70space} in conjunction with fixed-sized hash tables to consistently distribute flows to end hosts.
In rare occaisions we allow end hosts to drop some packets and use a controller to learn from the dropped packets to avoid future packet drops on the same connection.
Figure~\ref{fig0}~ surveys\footnote{In this paper we strictly refer to the utilization of service instances as their state. Service instances' state is different from the mapping between existing connections and assigned instances which is referred to as load balancers' state. In our terminology a stateful load balancer is one that monitors instances' state.} the existing works and compares our design choices to our peers.

Following summarizes our contributions:
\begin{itemize}
    \item Unlike existing works, INCAB requires no modifications at hosts' protocol stack; instead, it relies on a light-weight hypervisor service to monitor instances' utilization.
    \item INCAB frees memory resources at load balancers by using a Bloom filter. 
    Compared to Silkroad~\cite{miao2017silkroad} or Spotlight~\cite{aghdai2018spotlight}, INCAB has a much smaller memory footprint.
    \item INCAB does not rely traffic rerouting for PCC. We show that solutions that use traffic rerouting can add as much as 12.48\% of additional traffic to the network
    \item Our flow-level simulation sends more than one million flows to the load balancer and shows that INCAB improves connections' average flow completion time over stateless solutions (e.g., Beamer) by 31.97\%.
\end{itemize}

The rest of this paper is organized as follows.
\S~\ref{motivationSec} explains the advantages of in-network congestion-aware load balancing in more detail.
Next, we introduce INCAB and present its design from data plane and controller plane perspectives in \S\ref{dpSec} and \S\ref{cpSec}, respectively.
\S~\ref{evalSec} evaluates INCAB and compares its performance with existing solutions.
\S~\ref{relatedSec} surveys the related works in the area.
Finally, \S~\ref{conclusionSec} concludes the paper.

\section{Motivation and Background}\label{motivationSec}
In data centers, services are distinguished by their virtual IP addresses (VIPs).
A VIP typically has many instances that are uniquely identified by their direct IP addresses (DIPs).
Recent works proposed several mechanisms to maintain PCC on dynamic DIP pools.
For instance, \cite{olteanu2018stateless,araujo2018balancing,claudel2018stateless} leverage from the connection state that is stored at DIPs.
These solutions implement a stateless load balancer that hashes connections into entries that are mapped to DIPs.
If a DIP is added to or removed from the pool, the load balancer updates corresponding entries with new destinations from the DIP pool while memorizing the old mappings.
Connections that hit the modified entries are sent to the new destination.
As a result, the new destination receives \emph{all} of the traffic mapped to entries that include new flows and existing flows to the old DIP.
Since the new DIP has no state for existing flows, it forwards them to their old DIP which is tagged to each packet by the load balancer.
The traffic rerouting, also referred to as daisy chaining, guarantees PCC in such systems.

Given such advances, it may seem that adding an additional module to poll DIPs and include or exclude instances based on their utilization would be trivial. 
However, using a simple example we show the contrary, that is congestion aware load balancing is not trivial even using state-of-the-art DIP addition/removal techniques.

\begin{figure}[t]
    \begin{minipage}[t]{0.155\textwidth}
        \includegraphics[width=\textwidth]{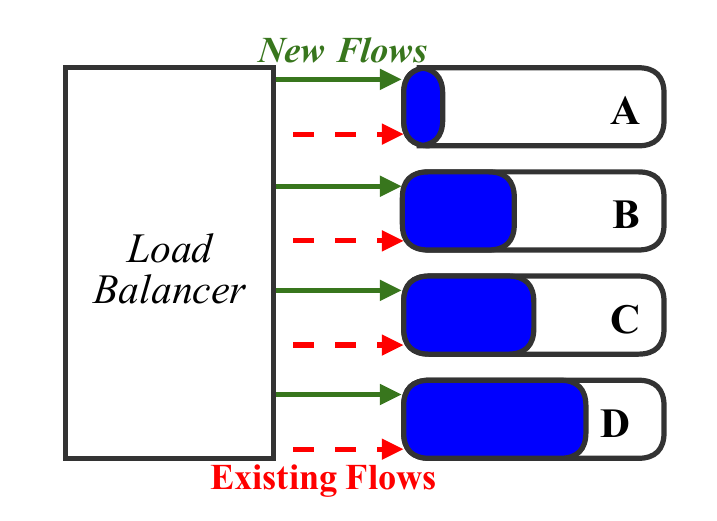}
        \subcaption{Stateless load balancing}
        \label{exampleA}
    \end{minipage}
    \begin{minipage}[t]{0.155\textwidth}
        \includegraphics[width=\linewidth]{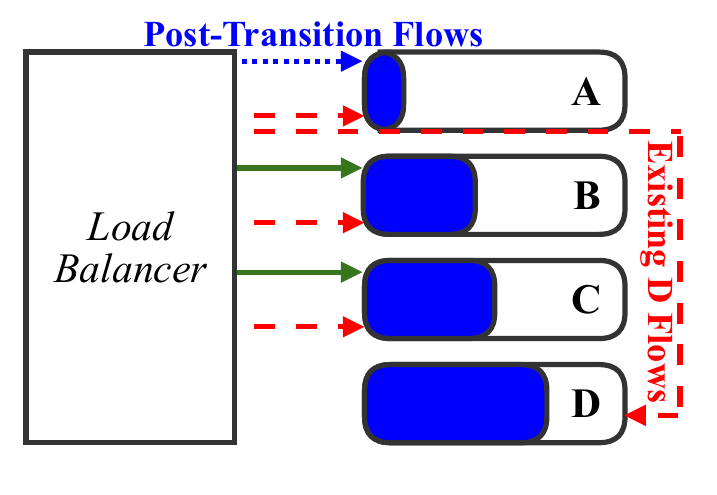}
        \subcaption{Traffic redirection}
        \label{exampleB}
    \end{minipage}
    \begin{minipage}[t]{0.155\textwidth}
        \includegraphics[width=\linewidth]{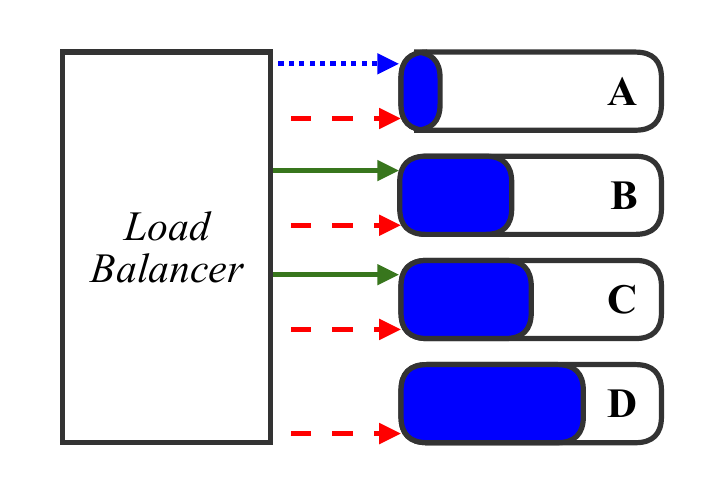}
        \subcaption{In-network load  balancing}
        \label{exampleC}
    \end{minipage}
        \caption{Load balancing methods}
\end{figure}

Load balancers use weights to assign incoming flows to DIPs; the set of DIP weights form the load balancing state.
Load balancing state may change due to a variety of reasons such as addition/removal of DIPs.
We refer to change in load balancing state as a transition from an old state to a new state.

Consider the example of Figure~\ref{exampleA} where a load balancer is distributing incoming traffic among four DIPs, one of which (\textbf{D}) is highly utilized, while another one (\textbf{A}) is under-utilized.
In this case, a congestion-aware load balancer gives a higher priority to \textbf{A} to alleviate the load imbalance.
Beamer provides a means to that goal.
As shown in Figure~\ref{exampleB}, the weight of \textbf{D} can be reduced while the weight of \textbf{A} is increased.
Weight changes are achieved by re-writing some hash table entries.
We refer to this change as a \emph{Trnasition} from \textbf{D} to \textbf{A}.
Daisy chaining sends \emph{all} of the traffic mapped to the updated hash table entries to \textbf{A}.
We make a distinction between new flows that arrive at unchanged destination (such as \textbf{B} and \textbf{C}) and destination in a transition (such as \textbf{A}); the latter are referred to as \emph{post-transition flows}.
In Beamer, \textbf{A} accepts post-transition flows and reroutes the existing flows to \textbf{D} to preserve PCC.
However, all of the flows to \textbf{A} and \textbf{D} share the same network bottleneck and have to compete for bandwidth.
As a result, new flows to \textbf{A} cannot fully utilize its resources and the rate of old flows to \textbf{D} may also be degraded.

This example shows that solutions that rely on traffic redirection at DIPs introduce new bottlenecks for the affected flows and may not be able to fully utilize DIP resources.
The bottleneck may not pose serious problems in a stateless load balancer due to the infrequent changes at the pool.
In a stateful load balancer, however, DIPs' share of new connections are dynamically adjusted during the run-time and daisy chaining is required at a much higher frequency as a result. 

An optimal solution, as shown in Figure~\ref{exampleC}, is for the load balancer to send flows assigned to \textbf{D} to the correct destination during the transition while sending transient flows to \textbf{A}.
Such a load balancer is referred to as in-network congestion-aware load Balancer (INCAB) and has the following properties:
\begin{enumerate}
    \item DIPs' share of new connections is adjustable at runtime for efficient load distribution.
    \item It avoids DIP-level traffic rerouting (e.g., daisy chaining) to minimize network resource overhead.
    \item Connection tracking is avoided to increase resilience and minimize the memory usage at load balancers.
\end{enumerate}

\section{INCAB Data Plane}\label{dpSec}
\begin{figure}[t]
    \centering
    \includegraphics[width=0.5\textwidth]{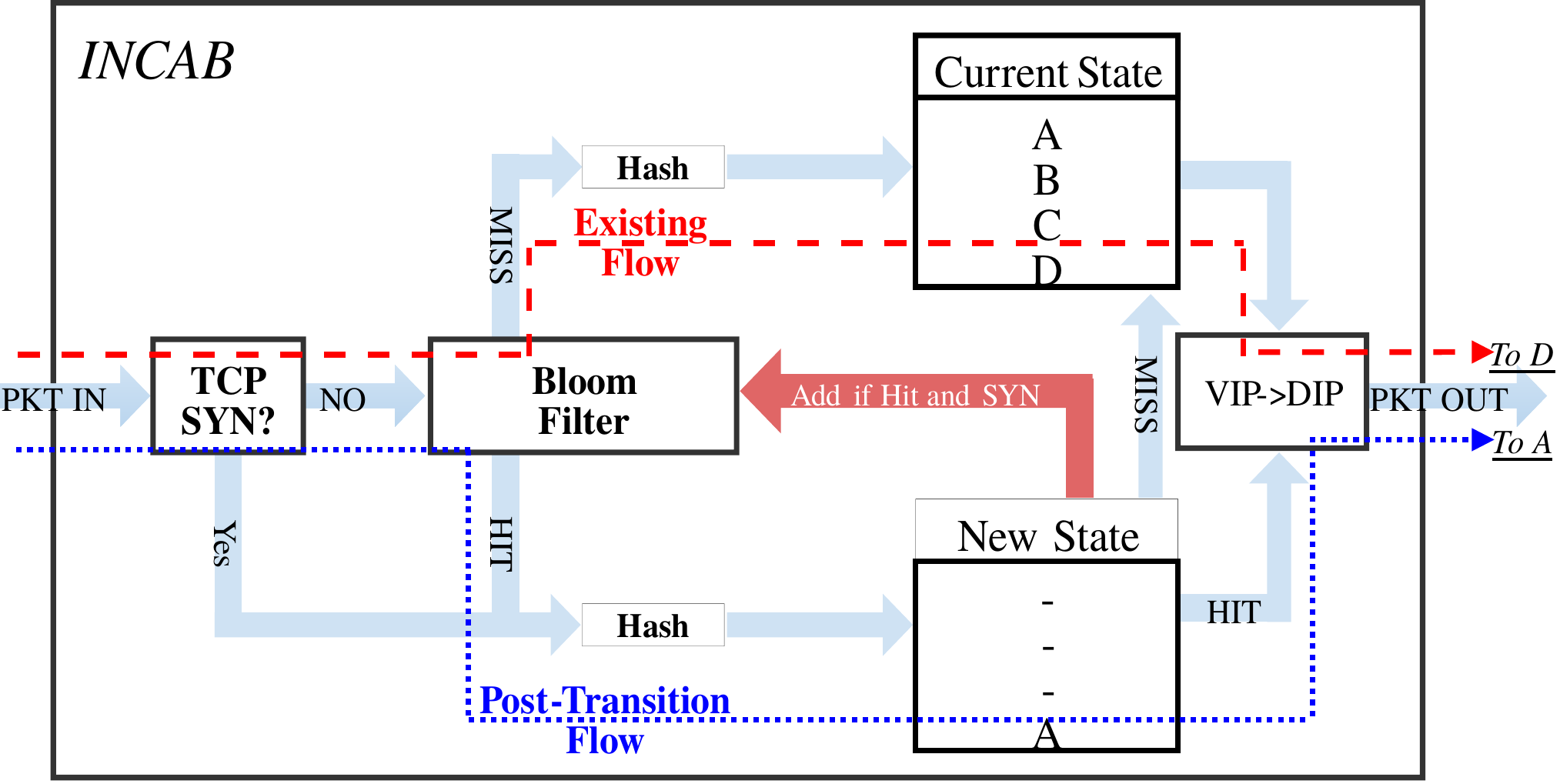}
    \caption{INCAB Architecture.}
    \label{detailDesign}
\end{figure}

Figure~\ref{detailDesign} shows INCAB's data plane.
INCAB utilizes hash tables with fixed number of entries mapped to the DIPs.
We use two different hash tables (with the same hash function) for existing flows and post-transition flows respectively.
In addition, it uses a Bloom filter to distinguish packets of post-transition flows from those of existing flows.

\begin{figure*}
    \centering
    \begin{minipage}[t]{0.235\textwidth}
        \includegraphics[width=\linewidth]{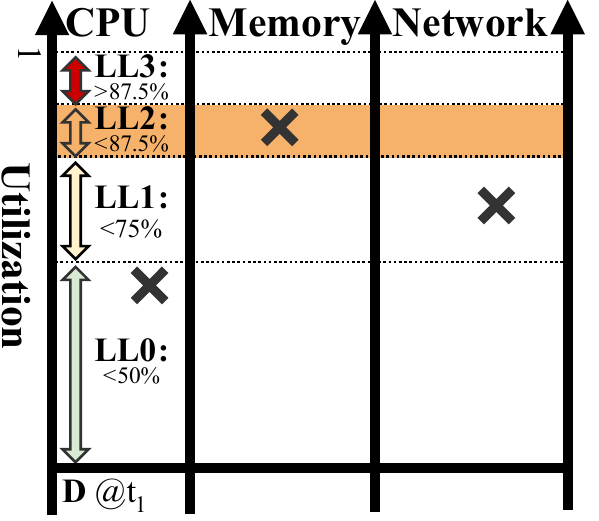}
        \caption{Highest-utilized resource determines LL.}
        \label{LLCalc}
    \end{minipage}
    \begin{minipage}[t]{0.235\textwidth}
        \includegraphics[width=\linewidth]{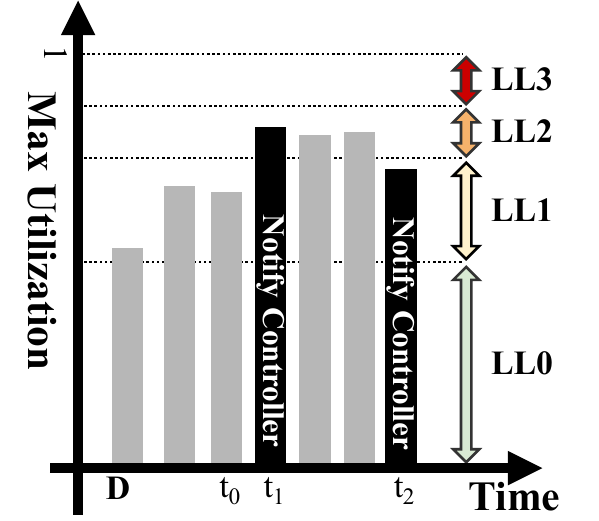}
        \caption{D notifies controller on LL changes.}
        \label{loadLevelFig}
    \end{minipage}
    \begin{minipage}[t]{0.235\textwidth}
        \includegraphics[width=\linewidth]{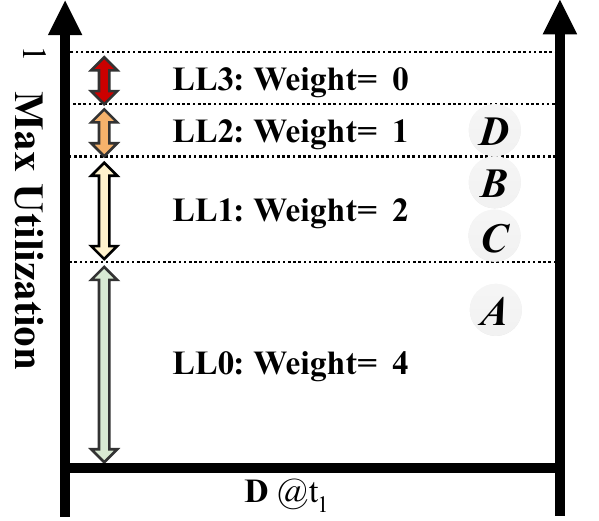}
        \caption{Low LLs get higher transition weights.}
        \label{LLExample}
    \end{minipage}
    \begin{minipage}[t]{0.25\textwidth}
        \includegraphics[width=\linewidth]{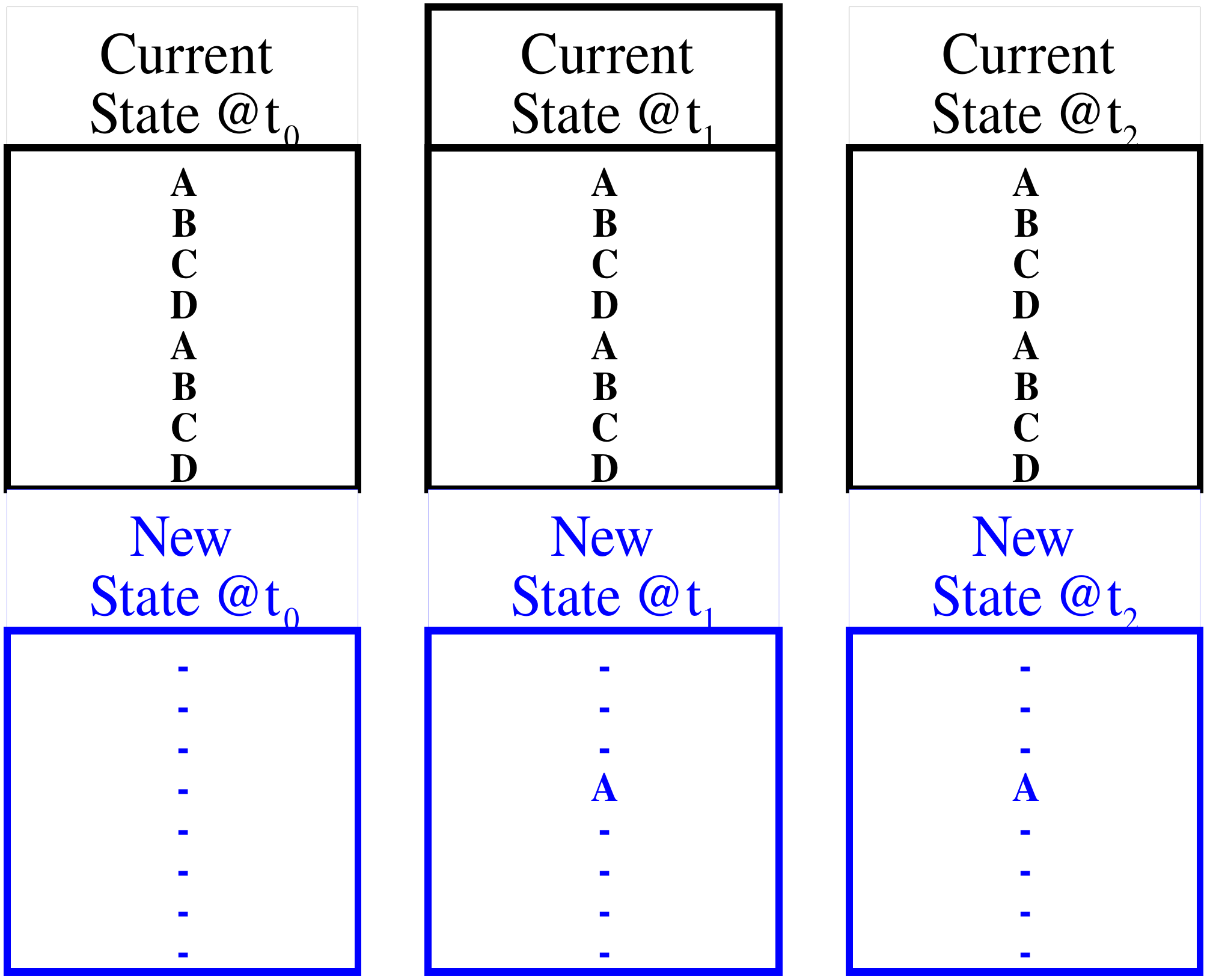}
        \caption{INCAB transitions in our example.}
        \label{transitionFig}
    \end{minipage}
\end{figure*}

To explain the operation of INCAB, we use different scenarios highlighted in different colors in Figure~\ref{detailDesign}.
Existing flows take the Red, dashed path where they merely mapped to a hash entry from the \emph{current state} of the system.
The number of hash table entries for each DIP in current state determine the relative weight of DIPs.
During transitions, we update the relative weight of DIPs and transition from the current state of DIP weights to a \emph{new state}.
Therefore, for DIPs in transition, we write some entries in new state hash table, while all of the corresponding hash table entries in new state for DIPs not in transition are empty.
INCAB guarantees PCC by ensuring that entries of current state are not modified as long as they have active flows.
We identify the first packet of new flows by monitoring TCP SYN flag and send them to the new state of the system.
If a new flow is mapped to an empty entry in the new state, it means that the flow is not post-transition.
Such new flows are sent to the current state hash table and their future packets will take the Red path.
The first packet of post-transition flows, however, is mapped to a non-empty entry of the new state.
INCAB adds 5-tuple identifier of post-transition flows to the Bloom filter to distinguish the rest of their packets from existing flows.
Figure~\ref{detailDesign} highlights the data path for post-transition flows using blue, dotted arrows.

The Bloom filter used for distinguishing post-transition flows may return a false positive hit, meaning that a flow may hit the Bloom filter without being added to it.
False positive rate depends on Bloom filter's size, number of hash functions, and number of occupants.
In Figure~\ref{detailDesign}, false positives force existing flows to be processed using new state.
INCAB can detect a subset of false positives.
In cases where the Bloom filter returns a false positive on a flow that is mapped to an empty entry of new state, INCAB detects the false positive and sends the flow back to current state hash table to preserve consistency.
In rare situations, Bloom filter false positives may happen to be mapped to a non-empty entry in new state in which case the architecture of Figure~\ref{detailDesign} breaks PCC.
The probability of such instances with a solution to maintain PCC is discussed in detail in \S~\ref{BFFPSec}.

\section{INCAB Control Plane}\label{cpSec}
Our load balancing heuristic adjusts DIPs' weights in a closed feedback loop.
Overwhelmed DIPs' weights are reduced while they still serve their existing connections.
This process is repeated until DIPs get approximately equal load.

The controller also monitors entries in current state that experience an idle time out.
A time-out event means that all existing connections to the corresponding entry are finished.
Therefore, the controller proceeds to rewrite the content of timed-out entry with that of its respective entry from the new state; then it empties the entry in new state.
The controller also removes post-transition flows of the timed-out entry from the Bloom filter.

\subsection{Congestion-Aware Transitions}
During transitions, INCAB routes post-transition flows using new state that point to new DIPs.
In normal operation, all DIPs have an equal number of entries on this table.
However, we can adjust the share of new flows for each DIP by changing the number of entries assigned to that DIP.

Instead of polling DIPs, INCAB relies on DIPs to notify the controller when they experience a change in their utilization.
DIPs have a variety of resources such as network bandwidth, CPU, memory, etc.
We define DIPs' \emph{load level (LL)} as the utilization level of their highest-utilized resource.
Figure~\ref{LLCalc} shows the utilization of \textbf{D}'s resources at $t_0$.
In this example, memory is the highest-utilized resource and it determines \textbf{D}'s LL.
DIPs that run the same application use the same resource type for load estimation to produce meaningful weights for load balancing.
Therefore determining the bottleneck resource requires us to study and analyze applications.
In situations where such information is not available from applications or in dynamic environments where various resource types can become the bottleneck at different times, the queue size (i.e., number of outstanding packets at DIPs' receive queue) can be used as a single metric that is affected by the highest-utilized resource regardless of its type.
INCAB defines a set of exponential thresholds, and rely on an agent at DIPs to send a notification to the controller when the LL is changed.
Figure~\ref{loadLevelFig} illustrates the load level of \textbf{D} over time and highlights the instances where INCAB controller is notified, e.g., at t1 and t2 as shown in the figure. 
DIPs use a sliding window to calculate the average load level.
INCAB agents slide the averaging window slightly (e.g., in $O(10ms)$) for smooth LL change at threshold boundaries.

INCAB controller's objective is to achieve fair load distribution by bringing all DIPs to the same LL.
INCAB controller uses announced LLs to determine DIPs' share in current/new state.
A heuristic algorithm at the controller periodically reduces the share of a DIP in the highest observed LL in half and initiates the transition of half of the entries of the highly-utilized DIP to other DIPs in the lower LLs.
As shown in Figure~\ref{LLExample} DIPs in lower levels have a higher chance of being chosen for transitions; each LL gets an exponential weight which corresponds to probability of choosing DIPs in that LL for future transitions.

In our example, we assume that \textbf{A} is at LL1, \textbf{B}, \textbf{C} are at LL2 , and the weight of A is 4 and that of B/D is 2 as shown in Figure~\ref{LLExample}.
As shown in Figure~\ref{transitionFig}, the controller start the transition on half of the entries belonging to \textbf{D} when its load reaches LL2 at $t_1$.
One entry is transitioned to \textbf{A}.
As shown in Figure~\ref{loadLevelFig}, at $t_2$ \textbf{D}'s LL drops to 1 which means that the highest observed LL is reduced to level 1.
The new state hash table is not changed in this case.
This heuristic algorithm continues to periodically initiate transitions until all DIPs reach the same LL.
As shown in Figure~\ref{transitionFig} the transition will not be terminated at $t_2$; new flows after $t_2$ are still considered as post-transition flows and will be routed to \textbf{A}.
To maintain PCC, transitions can only be terminated after \emph{all} existing flows to \textbf{D} end at which time we can rewrite the corresponding entry in current state with A and treat all active flows mapped to that entry as existing flows.
Termination of transitions is discussed in \S~\ref{terminationSubSec}.

\begin{figure}[b]
    \centering
    \includegraphics[width=0.5\textwidth]{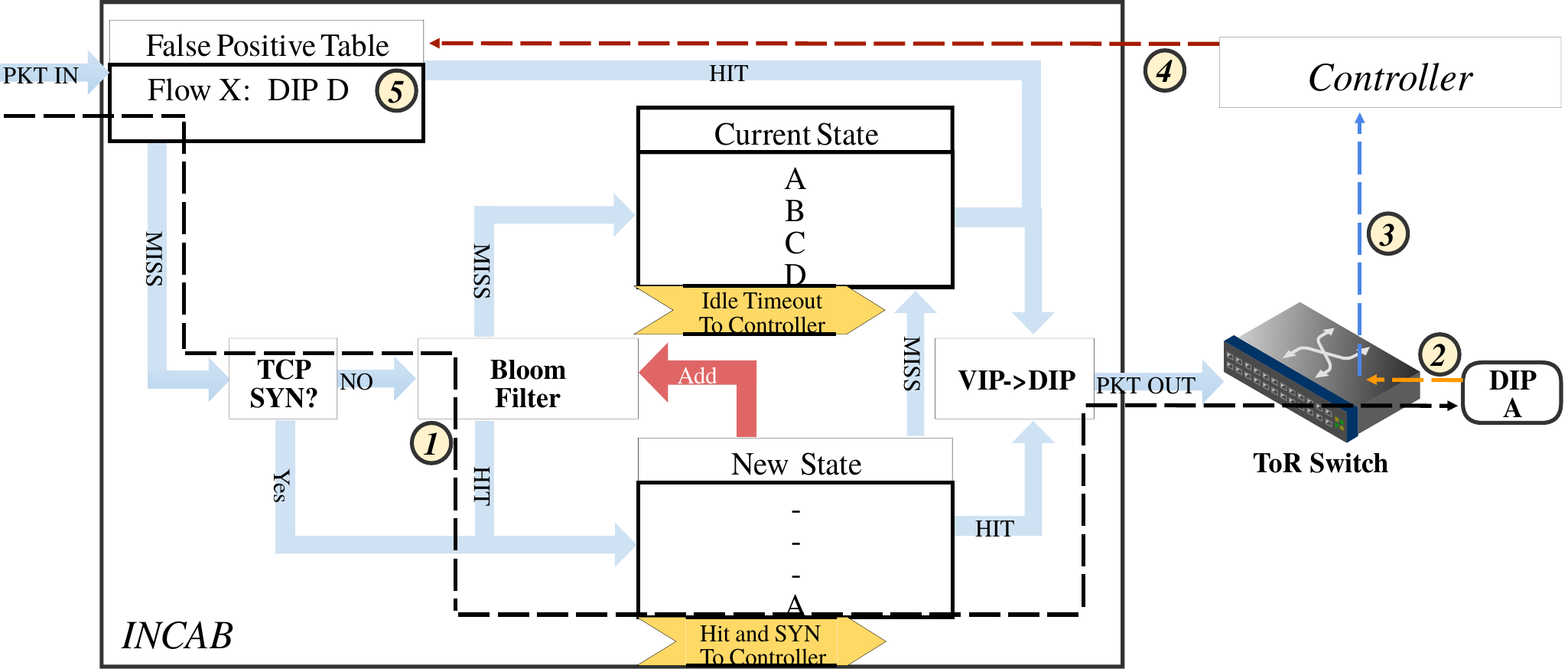}
    \caption{Tracking false positives to avoid PCC breaks.}
    \label{detailDesignTrack}
\end{figure}

\begin{figure*}[t]
    \centering
    \begin{minipage}[t]{0.3275\textwidth}
            \begin{tikzpicture}
        \begin{axis}[
                xlabel={MAWI Traffic Sample ID},
                ylabel={Normalized Average Flow Completion Time},
                major x tick style = transparent,
                ybar=2*\pgflinewidth,
                bar width=12pt,
                ymajorgrids = true,
                symbolic x coords={1,2,3},
		xtick = {data},
                scaled y ticks = false,
                enlarge x limits=0.50,
                ymin=0,
                ymax=1.1,
                legend cell align=left,
                legend columns=2,
                legend style={at={(0.5,1.11)},anchor=north},
            ]
            \addplot[style={fill=gray}]
            coordinates {
                (1, 1)
                (2, 1)
                (3, 1)
            };

            \addplot[style={fill=blue},pattern=north west lines]
            coordinates {
                (1,0.6635)
                (2,0.7104)
                (3,0.6670)
            };
            \legend{Stateless, INCAB}
        \end{axis}
\end{tikzpicture}
        \caption{Average completion time.}
        \label{figACT}
    \end{minipage}
    \begin{minipage}[t]{0.3275\textwidth}
            \begin{tikzpicture}
        \begin{axis}[
                xlabel={MAWI Traffic Sample ID},
                ylabel={Maximum False Positives Table Size},
                major x tick style = transparent,
                ybar=2*\pgflinewidth,
                bar width=12pt,
                ymajorgrids = true,
                symbolic x coords={1,2,3},
		xtick = {data},
                scaled y ticks = false,
                enlarge x limits=0.50,
                ymin=0,
                ymax=600,
                legend cell align=left,
		legend columns=2,
                legend style={at={(0.5,1.11)},anchor=north},
            ]

            \addplot[style={fill=blue},pattern=north west lines]
            coordinates {
                (1,521)
                (2,488)
                (3,594)
            };
            \legend{INCAB}

        \end{axis}
\end{tikzpicture}
        \caption{Max False positive table size.}
        \label{figFP}
    \end{minipage}
    \begin{minipage}[t]{0.3275\textwidth}
            \begin{tikzpicture}
        \begin{axis}[
                xlabel={MAWI Traffic Sample ID},
                ylabel={Percentage of Redirected Traffic},
                major x tick style = transparent,
                ybar=2*\pgflinewidth,
                bar width=12pt,
                ymajorgrids = true,
                symbolic x coords={1,2,3},
        xtick = {data},
                scaled y ticks = false,
                enlarge x limits=0.50,
                ymin=-0.1,
                ymax=15,
                legend cell align=left,
		legend columns=2,
                legend style={at={(0.5,1.11)},anchor=north},
            ]
            \addplot[style={fill=gray}]
            coordinates {
                (1, 12.98804)
                (2, 11.98963)
                (3, 12.46124)
            };
            \addplot[style={fill=blue},pattern=north west lines]
            coordinates {
                (1,0)
                (2,0)
                (3,0)
            };
            \legend{SHELL, INCAB}
        \end{axis}
\end{tikzpicture}
        \caption{Traffic redirection volume.}
        \label{figRedir}
    \end{minipage}
\end{figure*}

\subsection{Bloom Filter False Positives}\label{BFFPSec}

Assuming that hash functions select entries with equal probability for a Bloom filter with $k$ hash functions, size of $m$ entries, and $x$ inserted elements (i.e., number of entries in transition), probability of a positive rate is:
$(1 - e^{-kx/m})^k$.
Given that we are using N entries for hashing, only $x/N$ of Bloom filter false positives go through the transient entries and break PCC.
Hence, probability of a PCC break is:
\begin{equation}
    \frac{x(1 - e^{-kx/m})^k}{N}
    \label{fpEq}
\end{equation}

For instance, probability of PCC breaks for an INCAB with 64K entries, a Bloom filter of 256K entries with 2 hash functions and 1000 inserted flows is around 0.011\%.
We can estimate an average size for the false positive table using Eq. 1 and the average number of active flows at the load balancer.

We detect flows that break PCC and track them using an additional false positive table to meet PCC for all flows and eliminate the adverse impact of Bloom filter false positives.
This mechanism is shown in Figure~\ref{detailDesignTrack}.
In this example, the Bloom filter returns a false positive on an existing flow to \textbf{D} (1).
The flow is mapped to a non-empty entry in new state and is routed to \textbf{A}.
\textbf{A} receives the packet, and since it has no state for it, resets the connection with a TCP RST (2).
Top of the rack (ToR) switch routes TCP RSTs from IP range of DIPs to the controller (3).
RSTs' redirection can be realized using a simple OpenFlow rule on ToR.
The controller is aware of the transition from \textbf{D} to \textbf{A}, it rehashes the flow and finds its old target which is \textbf{D}.
Based on this information the controller adds a rule to the false positive table to bypass this particular flow and directly send it to \textbf{D} (4).
Future packets of this flow will hit the false positive table and are sent to \textbf{D} (5).
This mechanism eliminates false positives at a system level at cost of dropping a negligible number of packets.

INCAB obsoletes daisy chaining and meets PCC without modifying hosts' protocol stack. 
It tracks some connections; however, as we show in Eq.~\ref{fpEq}, its false positive rate is at least an order of magnitude smaller than a pure Bloom filter.
INCAB's false positive table does not consume as much resources as connection tables in existing works such as~\cite{aghdai2018spotlight,miao2017silkroad} due to a number of reasons:
\begin{enumerate}
    \item \emph{Not every flow is added to the false positive table.}
        The false positive table only admits some of the flows that are mapped to a hash table entry under transition.

    \item \emph{The usage of IDLE time-outs.}
        Data plane IDLE time outs can be used to phase out old entries in the hash table as well as entries in the false positive table.
        Therefore, our solution does not require controller involvement to erase the old entries in the hash table or the entries in the false positive table.

    \item \emph{False positive table size can be limited.}
        We can turn off polling or weight updates when a size-limited false positive table fully fills up.
        When the table fills up, the maintenance for addition/removals of DIPs can revert to DIP-level traffic redirection for meeting PCC.
\end{enumerate}

\subsection{Termination of Transitions}\label{terminationSubSec}
A transition is over when all of the existing flows that use the corresponding entry in current state finish.
INCAB uses a counting Bloom filter~\cite{fan2000summary} to enable flow removals once the transition is over.

INCAB sets an idle time-out on the entry of current state at the start of the transition.
We detect the end of transition when the idle time-out event triggers in which case the INCAB load balancer sends a notification to the controller as shown in Figure~\ref{detailDesignTrack}.
Once the controller receives the notification, it can safely rewrite the content of new state to current state and vacates the entry in new state.
INCAB also notifies the controller of every insertion to the Bloom filter and the controller removes all of the post-transition flows for finished transitions from the Bloom filter.

This mechanism ensures that INCAB always holds a minimum number of flows at the Bloom filter and reduces the probability of false positives according to Eq.~\ref{fpEq}.

We can also define a hard time-out for long flows that are mapped to a hash table entries under transition.
Load balancers notify the controller if flows hit such entries after the hard time-out.
The controller adds such flows to the false positive table. As a result, the future packets of such long flows will not hit the current state hash table, allowing the system to terminate the transitions deterministically.
Hard time-outs are different from idle time-outs; they work in tandem for termination of transitions.
The former triggers for flows that hit an entry in  after a long time-out and guarantees that transitions are terminated regardless of the duration of existing flows while the latter triggers for entries that are no longer hit by existing flows after a short time-out.

\section{Evaluation} \label{evalSec}
We have simulated a web service hosted by a content delivery network (CDN).
In the experiments, we simulate 1024 DIPs with heterogeneous servers.
The capacities of servers are 1Gbps in average.
Also, we assumed that requests are initiated by broadband users whose maximum bandwidth does not exceed 100Mbps.
In the simulation, flows are sampled from WIDE MAWI archive~\cite{cho2000traffic} and fed to the load balancer following the Poisson distribution.
The flow arrival rate is set as 80\% of the system capacity, which is around 5851 flows/second depending on the average flow size of the sampled flows.
The duration of each simulation is 5 minutes and the load balancer routes around 1,650,000 flows in each round.
In INCAB, the Bloom filter is assigned with two hash functions and a 64MB table.
To measure LL of each DIP, a sliding window with a width of 50ms is applied for calculating the moving average of DIP utilization.
The window slides 1ms each time.
Since 98\% of the values are the same after sliding, the utilization changes smoothly, and so does LL.
The smooth changes reduce the number of notifications that sent to the controller, because DIPs send a notification only when LL changes.
By doing so, we can react to the LL changes in a short time without continuously notifying the controller.
We consider the following performance metrics:
\begin{itemize}
    \item Load balancing efficiency which affects flows' average completion time.
        Figure~\ref{figACT} compares flows' average completion time between INCAB and stateless schemes (i.e., any solution that does not update DIP weights in run-time such as Beamer~\cite{olteanu2018stateless} and ECMP solutions~\cite{miao2017silkroad,gandhi2015duet}).
        INCAB reduces flows' average completion time by 31.97\%.
    \item The switch memory utilization in terms of size of INCAB false positive connection table.
        We measure the maximum size of false positive table. Over the course of simulations, the maximum size of this table, shown in Figure~\ref{figFP}, did not exceed 600.
        INCAB data plane detects more than 91\% of Bloom filter false positives (those that hit an empty entry in new state, see Figure~\ref{detailDesign}).
        Only 8.46\% of false positives hit a non-empty entry in new state; those flows were added to the false positive table.
    \item Pressure on networking fabric in terms of traffic redirection volume.
        We also measure how much traffic SHELL redirect back to the network due to the daisy chaining.
        Our results are illustrated in Figure~\ref{figRedir}.
        On average, SHELL roughly adds a 12.48\% traffic overhead to the aggregated volume of traffic served by the service.
\end{itemize}


\section{Related Works} \label{relatedSec}
ECMP is the most common technique for load balancing at transport layer.
Ananta~\cite{patel2013ananta} proposes softwarized ECMP-based load balancing for data centers.
Duet~\cite{gandhi2015duet} proposes hybrid load balancing; it keeps a connection table at software and uses commodity switches to perform ECMP-based flow dispatching.
Silkroad~\cite{miao2017silkroad} utilizes programmable switches to perform flow tracking as well load balancing at hardware.

Consistent hashing is another technique for flow dispatching.
Maglev~\cite{eisenbud2016maglev} uses consistent hashing for flow dispatching, however, in dynamic scenarios Maglev prioritizes efficient load balancing over maintaining PCC.
Faild~\cite{araujo2018balancing} targets CDN pods and it uses consistent hashing and DIP-level traffic redirection to implement unweighted L4 load balancing without using connection tables.
Beamer~\cite{olteanu2018stateless} implements a two-stage consistent hashing mechanism for flow dispatching at data center networks; it guarantees connection consistency through DIP traffic redirection  thout network state.

Congestion-aware load balancers take DIPs load into account during flow dispatching.
Spotlight~\cite{aghdai2018spotlight} polls instances' utilization and assigns DIP weights in proportion to their available capacities.
It uses DIP weights for flow dispatching and meets PCC by implementing a connection table for all flows.
6LB~\cite{yoann20186lb} uses IPV6 segment routing and power of 2 choices to reroute traffic to a randomly chosen second choice when the first choice is overwhelmed; this solution also requires in-network state and tracks all connections.
SHELL~\cite{claudel2018stateless} removes the network state from 6LB by using traffic redirection, but it may break PCC for some flows.


\section{Conclusion} \label{conclusionSec}
INCAB uses minimal network state for in-network congestion-aware load balancing.
It does not use connection tracking for every flow, thereby saving load balancers' memory.
It also avoids daisy chaining to save network bandwidth.

We use a flow-level simulation to evaluate INCAB's performance and compare it to existing works.
Our simulation follows a CDN network where a load balancer distributes clients' requests among a service cluster of 1024 DIPs.
We show that existing solutions that rely on traffic redirection may introduce as much as 12.48\% additional traffic to the service cluster.
Additionally, our results show that INCAB improves the average flow completion time by about 31.97\%.

INCAB achieves PCC using a Bloom filter and a compact false positive table.
We can estimate the average size of the false positive table using Eq. 1 and the average number of active flows. 
Our simulations show that with more than one million flows routed, the maximum size of the false positive table does not exceed 600.
Eq. 1 and the simulations results support the compact size of false positive table.
We aim to pursue further research on the following topics:
\begin{enumerate}
    \item \textbf{Implementation of INCAB using P4 language.}
        INCAB's data plane is compatible with the building blocks of P4 language;
        our peers implemented hash tables and Bloom filters using P4 ~\cite{miao2017silkroad}.
        We aim to build a small-scale testbed and implement INCAB using $P4_{16}$ and run its control plane using P4 runtime for further evaluation.
    \item \textbf{Formal analysis of control plane heuristics.}
        Our heuristic feedback algorithm shows significant potential in improving flows' average completion time;
        however, we plan to analyze its fairness using formal methods.
\end{enumerate}

\bibliographystyle{ieeetr}
\bibliography{references}

\begin{thebibliography}{10}

\bibitem{patel2013ananta}
P.~Patel, D.~Bansal, L.~Yuan, A.~Murthy, A.~G. Greenberg, D.~A. Maltz, R.~Kern,
  H.~Kumar, M.~Zikos, H.~Wu, C.~Kim, and N.~Karri, ``Ananta: cloud scale load
  balancing,'' in {\em {ACM} {SIGCOMM} 2013 Conference, SIGCOMM'13, Hong Kong,
  China, August 12-16, 2013}, pp.~207--218, 2013.

\bibitem{eisenbud2016maglev}
D.~E. Eisenbud, C.~Yi, C.~Contavalli, C.~Smith, R.~Kononov,
  E.~Mann{-}Hielscher, A.~Cilingiroglu, B.~Cheyney, W.~Shang, and J.~D. Hosein,
  ``Maglev: {A} fast and reliable software network load balancer,'' in {\em
  13th {USENIX} Symposium on Networked Systems Design and Implementation,
  {NSDI} 2016, Santa Clara, CA, USA, March 16-18, 2016}, pp.~523--535, 2016.

\bibitem{miao2017silkroad}
R.~Miao, H.~Zeng, C.~Kim, J.~Lee, and M.~Yu, ``Silkroad: Making stateful
  layer-4 load balancing fast and cheap using switching asics,'' in {\em
  Proceedings of the Conference of the {ACM} Special Interest Group on Data
  Communication, {SIGCOMM} 2017, Los Angeles, CA, USA, August 21-25, 2017},
  pp.~15--28, 2017.

\bibitem{olteanu2018stateless}
V.~A. Olteanu, A.~Agache, A.~Voinescu, and C.~Raiciu, ``Stateless datacenter
  load-balancing with beamer,'' in {\em 15th {USENIX} Symposium on Networked
  Systems Design and Implementation, {NSDI} 2018, Renton, WA, USA, April 9-11,
  2018}, pp.~125--139, 2018.

\bibitem{araujo2018balancing}
J.~T. Ara{\'{u}}jo, L.~Saino, L.~Buytenhek, and R.~Landa, ``Balancing on the
  edge: Transport affinity without network state,'' in {\em 15th {USENIX}
  Symposium on Networked Systems Design and Implementation, {NSDI} 2018,
  Renton, WA, USA, April 9-11, 2018}, pp.~111--124, 2018.

\bibitem{aghdai2018transparent}
A.~Aghdai, M.~Huang, D.~Dai, Y.~Xu, and J.~Chao, ``Transparent edge gateway for
  mobile networks,'' in {\em 2018 IEEE 26th International Conference on Network
  Protocols (ICNP)}, pp.~412--417, IEEE, Sept 2018.

\bibitem{aghdai2019enabling}
A.~Aghdai, Y.~Xu, M.~Huang, D.~H. Dai, and H.~J. Chao, ``Enabling mobility in
  lte-compatible mobile-edge computing with programmable switches,'' {\em arXiv
  preprint arXiv:1905.05258}, 2019.

\bibitem{gandhi2015duet}
R.~Gandhi, H.~H. Liu, Y.~C. Hu, G.~Lu, J.~Padhye, L.~Yuan, and M.~Zhang,
  ``Duet: Cloud scale load balancing with hardware and software,'' {\em ACM
  SIGCOMM Computer Communication Review}, vol.~44, no.~4, pp.~27--38, 2015.

\bibitem{aghdai2018spotlight}
A.~Aghdai, C.~Chu, Y.~Xu, D.~H. Dai, J.~Xu, and H.~J. Chao, ``Spotlight:
  Scalable transport layer load balancing for data center networks,'' {\em
  CoRR}, vol.~abs/1806.08455, 2018.

\bibitem{yoann20186lb}
Y.~Desmouceaux, P.~Pfister, J.~Tollet, M.~Townsley, and T.~H. Clausen, ``6lb:
  Scalable and application-aware load balancing with segment routing,'' {\em
  {IEEE/ACM} Trans. Netw.}, vol.~26, no.~2, pp.~819--834, 2018.

\bibitem{claudel2018stateless}
B.~Pit-Claudel, Y.~Desmouceaux, P.~Pfister, M.~Townsley, and T.~Clausen,
  ``Stateless load-aware load balancing in p4,'' in {\em 2018 IEEE 26th
  International Conference on Network Protocols (ICNP)}, pp.~418--423, IEEE,
  Sept 2018.

\bibitem{karger1997consistent}
D.~R. Karger, E.~Lehman, F.~T. Leighton, R.~Panigrahy, M.~S. Levine, and
  D.~Lewin, ``Consistent hashing and random trees: Distributed caching
  protocols for relieving hot spots on the world wide web,'' in {\em
  Proceedings of the Twenty-Ninth Annual {ACM} Symposium on the Theory of
  Computing, El Paso, Texas, USA, May 4-6, 1997}, pp.~654--663, 1997.

\bibitem{benson2010network}
T.~Benson, A.~Akella, and D.~A. Maltz, ``Network traffic characteristics of
  data centers in the wild,'' in {\em Proceedings of the 10th {ACM} {SIGCOMM}
  Internet Measurement Conference, {IMC} 2010, Melbourne, Australia - November
  1-3, 2010}, pp.~267--280, 2010.

\bibitem{greenberg2009vl2}
A.~G. Greenberg, J.~R. Hamilton, N.~Jain, S.~Kandula, C.~Kim, P.~Lahiri, D.~A.
  Maltz, P.~Patel, and S.~Sengupta, ``{VL2:} a scalable and flexible data
  center network,'' in {\em Proceedings of the {ACM} {SIGCOMM} 2009 Conference
  on Applications, Technologies, Architectures, and Protocols for Computer
  Communications, Barcelona, Spain, August 16-21, 2009}, pp.~51--62, 2009.

\bibitem{aghdai2013traffic}
A.~Aghdai, F.~Zhang, N.~Dasanayake, K.~Xi, and H.~J. Chao, ``Traffic
  measurement and analysis in an organic enterprise data center,'' in {\em High
  Performance Switching and Routing (HPSR), 2013 IEEE 14th International
  Conference on}, pp.~49--55, IEEE, 2013.

\bibitem{bloom70space}
B.~H. Bloom, ``Space/time trade-offs in hash coding with allowable errors,''
  {\em Commun. {ACM}}, vol.~13, no.~7, pp.~422--426, 1970.

\bibitem{fan2000summary}
L.~Fan, P.~Cao, J.~M. Almeida, and A.~Z. Broder, ``Summary cache: a scalable
  wide-area web cache sharing protocol,'' {\em {IEEE/ACM} Trans. Netw.},
  vol.~8, no.~3, pp.~281--293, 2000.

\bibitem{cho2000traffic}
K.~Cho, K.~Mitsuya, and A.~Kato, ``Traffic data repository at the {WIDE}
  project,'' in {\em Proceedings of the Freenix Track: 2000 {USENIX} Annual
  Technical Conference, June 18-23, 2000, San Diego, CA, {USA}}, pp.~263--270,
  2000.

\end{thebibliography}

\end{document}